\documentstyle[preprint,revtex]{aps}

\newcommand{\eqn}[2]{
\begin{equation}
\label{#1} #2
\end{equation}
}

\newcommand{\eqna}[1]{
\begin{eqnarray} #1
\end{eqnarray}
}

\begin{document}
\pagestyle{empty}
\rightline{UNITUE-THEP-6-1993  \hfill May 13, 1993. }
\vspace{.5cm}
\begin{title}
\large
Anisotropic electron coupling as a phenomenological model for \\
high-temperature superconductors
\end{title}
\vspace{1cm}
\author{K. Langfeld$^{1)}$, E. Frey$^{2) 3)}$}
\begin{instit}
$^{1)}$ Institut f\"ur Theoretische Physik, Unversit\"at T\"ubingen,
D--7400 T\"ubingen, FRG
and
$^{2)}$ Physik-Department, Technische Universit\"at M\"unchen,
D--8046 Garching, FRG
$^{3)}$ Lyman Laboratory of Physics, Harvard University,
Cambridge, MA 02138, USA
\end{instit}
\begin{abstract}

A three-dimensional weak coupling BCS model with an {\it anisotropic}
pairing interaction in momentum space is reported. It exhibits an anisotropic
gap in accord with recent experimental observations for high-$T_c$ oxides.
The gap ratio  ${2 \Delta  / k_B T_c }$ is calculated as function of the
anisotropy ratio of the electron pairing. It is found that the mean gap ratio
decreases with increasing anisotropy of the electron pairing, whereas the
maximum gap ratio increases. Assuming a unique strength of the
effective electron interaction the transition temperature turns out to
be always larger than in the isotropic case.
The temperature dependence of the gap in the Cu-O plane and along the c-axis
of YBaCuO is presented.

\end{abstract}

$^{2)}$ Present address: TU Munich
\vskip 12pt

\begin{narrowtext}
\newpage
\pagestyle{plain}
\pagenumbering{arabic}

Since the discovery of the high temperature
copper-oxide superconductors~\cite{1,2,3} there has been great activity
in understanding the underlying microscopic mechanism of
superconductivity in these materials. Despite of great efforts
the fundamental pairing mechanism and the magnitude and symmetry of the
order parameter (gap) are still open questions. In this paper, we study
a BCS-like pairing theory with an {\it anisotropic} attractive
interaction between the quasiparticles.

The crystal structure of many of the high temperature superconductors features
Cu-O planes leading to unusually anisotropic superconducting properties.
A common feature is the nearly two-dimensional motion of the electrons
within the Cu-O planes~\cite{4}.
This is also reflected in the very large anisotropy observed in the
normal state resistivity tensor~\cite{5}, suggesting that electron
scattering is enhanced if the final momentum of the scattered
electron lies within the Cu-O planes. Furthermore
the broad range of values for
the ratio $2 \Delta / k_B T_c$ found in far-infrared
reflectivity~\cite{6,7,8,9,10} ($0<2 \Delta / k_B T_c < 8$) and tunneling
experiments~\cite{11,12} ($4.0<2 \Delta / k_B T_c < 7.6$) may indicate a
highly anisotropic gap. These values for the gap ratio seem to be beyond
the scope of weak coupling BCS theory~\cite{13} which yields
$r = 2 \Delta / k_B T_c = 3.5$ for an isotropic pairing interaction.
In the highly anisotropic BiSrCaCuO
even values of the order $r = 10$ have been measured~\cite{14}. Recently
there has been experimental evidence for a strong anisotropy
of the superconducting gap of the high-$T_c$ oxides~\cite{14,15,16,17,18,19}.
For example in NdCeCuO $2 \Delta $
in the Cu-O plane is $8\,$meV~\cite{18} whereas $6.25\, $meV
was measured for $2 \Delta $ in c-axis direction. In
TlBaCaCuO the ratio of the maximal to the minimal gap
$\Delta_{\rm max} / \Delta _{\rm min}$ is even about 6~\cite{15}.

Most of the theories proposed for HTC compounds emphasize their
two-dimensionality. BCS-pairing theories with interlayer charge
transfer excitations as the intermediate boson~\cite{20}
have been proposed leading
to an isotropic gap with a gap ratio $2 \Delta / k_B T_c \approx 10 ... 14$.
Recently, phenomenological models for layered superconductors, where the
quasiparticles interact via BCS-like pairing in the Cu-O planes, have been
proposed by several authors~\cite{21}.
The basic feature of those models is that in
addition to the dominant intralayer coupling a weak interlayer coupling is
introduced, which accounts for interlayer hopping processes including Josephson
tunneling. Main results are that for a pure intralayer pairing mechanism
the gap can not exhibit an anisotropy between the c-axis and the a-b-plane,
even
when one allows for single particle interlayer hopping. It is found to be
essential
that there is a spatial separation between the paired quasiparticles.
Furthermore, those models show that the existence of an anisotropic gap depends
on certain microscopic details like the number of conducting layers per unit
cell edge.

Here we disregard the discrete layered structure and adapt a more coarse
grained picture, where the pairing between the quasiparticles is
described by a  three-dimensional weak coupling BCS-model with an anisotropic
attractive interaction in momentum space.
Our main aim is to present a simple model which resembles the
universal features of a wide class of high temperature superconductors.
One objective of our work is to show that one can have a large gap ratio
$2 \Delta / k_B T_c$ even within the framework of
a weak coupling BCS theory if one accounts
for the anisotropy in the quasiparticle pairing.
We present results of the model
calculation for the gap ratio $2 \Delta _{\rm max} / k_B T_c$ and the
superconducting gap as a functional of the anisotropy of the electron
pairing and of temperature and compare them with
experiments.

In order to motivate our approach we start from a free electron Hamiltonian
with a linear coupling to phonons or some other intermediate bosons
\eqna{H \; =
      &&\; \sum_{p,\alpha=\pm} \xi_p b^{\dagger }_{p\alpha }
        b_{p\alpha } \; + \;
        \sum_{p} \omega_p a^{\dagger }_{p } a_{p } \nonumber \\
    + &&\; \sum_{p q \alpha } g( {\hat c} {\hat p} )
         [ b^{\dagger }_{p \, \alpha } b_{p+q \, \alpha } a^{\dagger }_{q} \, +
\,
           b^{\dagger }_{p+q \, \alpha } b_{p \, \alpha } a_{q} ] \; . }
Here $b^{\dagger }_{p\alpha }$ and $a^{\dagger }_{p }$ are
creation operators and
$\xi_p=p^2/2m-\mu$ and $\omega_p$
are one-particle energies for electrons and phonons respectively.
We study both a spherical Fermi sphere with
$p^{2}=p_{x}^{2}+p_{y}^{2}+p_{z}^{2}$ and with $p^{2}=p_{x}^{2}+p_{y}^{2}$,
$p_{z} \| \hat{c}$ a cylindric Fermi surface.
In a more realistic calculation one would have to consider
a more or less complicated Fermi surface. One might
expect that anisotropies of the Fermi surface are less important for
the superconducting phase transition, since it is well known that
standard BCS models with anisotropic Fermi surface but an isotropic
pairing coupling result in an isotropic gap function~\cite{21}.
Actually we will show later in the text that the weak coupling gap equation
is the same for cylindric and spherical Fermi surfaces.
This gives us a further hint, that the key point
for an anisotropic gap is an anisotropic (attractive) interaction between the
quasiparticles.
In order to account for the crystal medium we have introduced a direction
dependent electron-phonon coupling constant $g( {\hat c} {\hat p} )$,
where $\hat c$ and $\hat p$ are unit vectors, and $\hat c$ points
towards the c-axis direction of the crystal. The function
$g( {\hat c} {\hat p} )$ is assumed to be maximal for
${\hat c} {\hat p} = 0$.  This choice ensures that the electron-phonon
scattering amplitude is enhanced if the final momentum of the electron
lies within the CuO-plane. Especially, this implies that the
properties of the real crystal medium
are included into our free Hamiltonian.

The electron-phonon coupling provides an electron-electron scattering
amplitude in second order perturbation theory. In order to eliminate
the phonons from the theory we replace the electron-phonon
interaction by a two particle electron operator which yields the same
amplitude in first order~\cite{13}.
Assuming that the phonon interaction
is short ranged (as in usual BCS theory) we obtain an
effective Hamiltonian for the
electrons
\eqna{ H \; =
      &&\; \sum_{p,\alpha=\pm} \xi_p b^{\dagger }_{p\alpha } b_{p\alpha }
\; \nonumber \\
    - &&\; \Delta^{-1}
        \sum_{p q k \alpha \beta } g( {\hat c} {\hat p} )
        g( {\hat c} {\hat k} ) b^{\dagger }_{p \alpha } b_{p+q \, \alpha }
        b^{\dagger }_{k+q \, \beta } b_{k \, \beta } \; ,}
where $\Delta $ is the phonon propagator in the short range approximation.
In the following we would like to focus on the pairing mechanism
of electrons. We therefore continue with the reduced Hamiltonian
corresponding to s-wave pairing
\eqn{3}{
H \; = \; \sum_{p,\alpha=\pm} \xi_p b^{\dagger }_{p\alpha }
b_{p\alpha } \; + \; \sum_{p,k} b^{\dagger }_{k+} b^{\dagger }_{-k-}
\, W_{kp} \, b_{-p-} b_{p+} \; , }
The interaction matrix has a factorized form
\eqn{4}{
W_{kp} \, := \, G \, f( {\hat c}  {\hat k}) \,
f( {\hat c}  {\hat p}) \hbox to 1 true cm{\hfil } \, , }
where $f$ is proportional to $g$.
Since the nature of the attractive interaction and
the type of Cooper pairing in
high temperature superconductors are still lively discussed, we do not
attempt to derive the above anisotropic interaction rigorous way.
Rather, we focus on investigating
the consequences of such an anisotropic pairing.
Since the dominant pairing interaction is within the Cu-O plane the anisotropy
function $f(x)$ will be peaked at small values of $x$. We choose $f$
to be normalized according to
\eqn{5}{
<f> \; := \; {1 \over 2} \int_{0}^{\pi} d\theta \; \sin \theta  \;
f( \cos \theta ) \; = \; 1 \; .}
so that the total coupling strength is given by an explicit factor $G$.

A model of this type was first discussed by D.\ Markowitz and L.\ P.\
Kadanoff~\cite{22}
in 1963. They  found that the anisotropy function was close to unity
$( <(f-1)^2> \, < \, 0.01 )$ for the metallic superconductors $Sn$ and
$Al$. Note that in the limit $f(x)\approx 1$ the usual isotropic BCS theory is
recovered from (3) and (4). A.\ J.\ Benett subsequently calculated the
superconducting properties of $Pb$~\cite{23} and reproduced the ratio
$2 \Delta / k_BT_c \approx 4$ from the experiment.

Whereas in those low-$T_c$ materials the anisotropy
is quite small and constitutes
only a small perturbation, high $T_c$ superconductors are almost
two-dimensional. Therefore, the anisotropy can no longer be treated as a
perturbation!

We proceed by solving the model (3) in the mean field approximation
neglecting fluctuations. The basic gap equation can be derived by standard
techniques~\cite{24}
\eqn{6}{
\Delta_k \; = \; {1 \over 2} \sum_{p} W_{kp} \,
{\Delta _{p} \over \sqrt{ \Delta_p^2 + \xi_p^2 } } \,
\hbox{tanh}( { \sqrt{ \Delta_p^2 + \xi_p^2 } \over 2 k_BT } ) \; . }
Introducing the interaction matrix from Eq.~(4) and inserting the Ansatz
\eqn{7}{
\Delta _k \; = \; f( {\hat c}  {\hat k} ) \; \Delta }
we obtain for spherical and cylindric Fermi surface a unique
result, i.\ e.
\eqna{
\Delta \; = &&\; g \int_{-1}^{1} dx \; \int_{-k\Theta_D}^{k\Theta_D} d\xi \;
{\Delta f^2(x) \over \sqrt{ \Delta^2 f^2(x) + \xi^2 } } \, \nonumber \\
&&\times \hbox{tanh}( { \sqrt{ \Delta^2 f^2(x) + \xi^2 } \over 2 k_BT } ) \; ,
}
where $\Theta _D$ is the Debye temperature and $k_B\Theta_D \ll \mu $
is assumed to arrive at (8).
Here $g$ is a numerical factor depending on the geometry of the
Fermi surface, and is proportional to $G$, which is the only parameter
in the theory. We choose $G$ to reproduce
the correct value for $\Delta $ at zero temperature.
Note, that the weak coupling BCS
assumptions yield a gap equation which is independent of the details
of the Fermi surface.
With (7) $\Delta $ can be interpreted as the gap average over all directions.
We remark
that the Debye temperature usually also depends on $x$. However,
in the physical region $k_B T \ll k_B\Theta _D \ll \mu $ the solution
of (8) is independent of $\Theta _D$ and it is allowed to set
the Debye temperature constant.
At the transition temperature equation (8) becomes
\eqn{9}{
1 \; = \; 2 g <f^2> \int {d\xi \over \xi }
\; \hbox{tanh}( {\xi \over 2 kT} ) \, , }
which is of the same structure as in the usual BCS theory~\cite{13,24}.
Hence we obtain
\eqn{10}{
T_c \; = \; {2 e^\gamma \over \pi } \, \Theta_D \, \exp( - {1 \over
<f^2> g } ) \; }
for the transition temperature, where $\gamma$ is Euler's constant.
In order to compare $T_{c}$ of the superconducting oxides to the ones
of the isotropic case we assume that $\Theta_D$ and the average
strength of the effective electron interaction $g$ are always in the
same order of magnitude. In this case the inequality
$<f^2> \geq <f>^2 =1 $, which holds for arbitrary functions $f$,
directly implies a larger $T_{c}$ for the HTC compounds. The transition
temperature is limited by $\Theta_D$. Note, however, that (10) is
only valid for $T_{c} \ll \Theta_D$ since a large $T_c$ implies
a large maximum gap (see below) which would contradict the weak coupling
assumption.

The physical value of $g$ is obtained by fixing the gap at zero
temperature
\eqn{11}{
{1 \over g} \; = \; \int dx \; d\xi \; { f^2(x) \over
\sqrt{ f^2 \Delta_0^2 + \xi ^2 } } \; \approx
\int_{-1}^{1} dx \; f^2(x) \ln { 2 \Theta _D \over f(x) \Delta_0 } \; , }
where $f \Delta_0 \ll k_B\Theta_D$ is assumed. Inserting Eq.~(11) into Eq.~(10)
our final result for the mean gap ratio is
\eqn{12}{
{2 \Delta _0 \over k_BT_c } \; = \; ({2 \Delta _0 \over k_BT_c } )_{\rm BCS}
\; \exp \{ - {\int dx \; f^2 \, \ln f \over \int dx \; f^2 } \} \; . }
The anisotropy function $f$ is directly related to the anisotropic
gap by Eq.~(7) and is therefore easily measured in the experiment.
Note,
that it can also be extracted from the coherence length which is
inverse proportional to the gap. It is important to note
that only the anisotropy function $f$ and the
average gap at zero temperature $\Delta _0$ is needed to predict the
transition temperature from Eq.~(12).

The task of a microscopic theory would be to calculate the anisotropy
function $f(x)$ from first principles.
Because of the lack of a precise experimental and theoretical knowledge of the
anisotropy function we proceed by investigating
a certain class of anisotropy functions
\eqn{13}{
f_\nu (x) \; = \; {\cal N} \, \exp ( - \mid { x \over \delta } \mid ^{\nu } )
 }
with two parameters $\nu, \delta $ to fit the anisotropy of the
electron pairing. Other assumptions about the anisotropy function can
be found in~\cite{21} where intra- and interlayer pairing models
have been considered.
At fixed $\nu $ the parameter $\delta $ is directly related to the
maximum of the gap anisotropy $\Delta _{\rm max} / \Delta _{\rm min} =
f_\nu (0) / f_\nu (1)$. Fig.\ 1 shows the anisotropy function and
the direction dependent gap for $\nu=1.2$ and $\Delta _{\rm max}
/ \Delta_{\rm min} =4.9$, which is a realistic ratio for YBaCuO~\cite{16,17}.
With the choice (13) and $\nu=1.5$ the quantity $2 \Delta / k_BT_c$, for the
mean gap $\Delta = \Delta_0$ and the maximum gap $\Delta = \Delta_{\rm max}$
was calculated as a function of the gap anisotropy.
The result is shown in Fig.\ 2. The average
gap ratio $r = 2 \Delta_0 / k_B T$ decreases with increasing anisotropy
whereas the maximum gap ratio
approaches a constant value. In the limit $\Delta_{\rm max} /
\Delta_{\rm min} \rightarrow \infty$ $(\delta \rightarrow 0)$ for a given class
$\nu $ the integrals in (12) can be solved analytically. The
asymptotic value is
\eqn{14}{
({2 \Delta _{\rm max} \over k_BT_c } )_{\rm asy} \; = \;
({2 \Delta _0 \over k_BT_c } )_{\rm BCS} \; \exp ( {1 \over 2 \nu } ) \; . }
For $\nu=1.5$ the above gap ratio is about 4.9, which is an average value
for a wide class of superconducting oxides~\cite{15}. Unfortunately
the gap ratio $\Delta_{\rm max} / \Delta_{\rm min}$ is measured only
for a few oxides. The experimental values shown in Fig.\ 2 are
taken from refs.~\cite{16,17} for YBaCuO, from refs.~\cite{18,19} for NdCeCuO
and from ref.~\cite{15}
for TlBaCaCuO. For testing the ideas it is presumable convenient to
investigate TlBaCuO compounds. It is observed that by controlling
the amount of oxygen in these compounds the gap keeps constant
whereas the transition temperature varies from 49 K to 120 K~\cite{25}.
In the framework of our model this would correspond to an increase in the
anisotropy of the electron-electron interaction.
Unfortunately, the anisotropy function $f$ or at least the
gap ratio is not yet measured. For $\nu =1.2$ the experimental data
for YBaCuO~\cite{16,17} are reproduced namely ${2 \Delta _{\rm max} / k_BT_c }
\approx 5$ at a gap ratio of 4.

Fig.\ 3 shows the temperature
dependence of the gap in the Cu-O plane ($\Delta_{\rm max}$)
and of the c-axis gap ($\Delta_{\rm min}$) calculated from (8) and (7).
For transition temperatures near $T_c$ the gap vanishes proportional
to $\sqrt{1-T/T_c}$, which is the usual result for a mean field
calculation. Deviations from this behavior are recently found
for YBaCuO~\cite{26}, which are probably due to fluctuation effects.
Note that BiSrCaCuO shows the mean field dependence on $T$ and
the qualitative behavior of Fig.\ 3 for the two gaps~\cite{14}.
BiSrCaCuO, however,  seems to be an almost two-dimensional superconductor
since a very small value of $\nu $ is needed to reproduce
${2 \Delta _{\rm max} / k_B T_c } \approx 10$ at a small gap ratio
$\Delta_{\rm max} / \Delta_{\rm min} \approx 1.3$.
Since these compounds have a coherence length in the c-direction which is much
smaller than the distance between the Cu-O layers, the discreteness of the
structure becomes important and an anisotropic model is no longer appropriate.

This class of
superconducting oxides seems to be beyond our simple model and one
has to apply are more detailed prescription similar as it was done in
refs.~\cite{20,21}.

In conclusion we have presented a
three-dimensional {\it weak coupling BCS model}
with an {\it anisotropic pairing} which correctly reproduces
the large value of ${2 \Delta _{\rm max} / k_BT_c } \approx 5$
for a wide class of superconducting oxides.
This result is obtained for both spherical and cylindrical Fermi surface.
We suppose therefrom that there is only a weak dependence of the
results on the precise form of the Fermi surface.
The anisotropy function $f$ can be directly extracted from the gap anisotropy
$\Delta (\theta ) =  <\Delta> f(x=\cos \theta)$, where
$\theta $ is the angle relative to the c-axis. We have presented a formula
which allows to  calculate $T_c$ solely from the properties of the
superconducting  oxide at zero temperature.
Assuming a unique order of magnitude for $\Theta _D$ and the strength of the
average electron interaction the model explains the large transition
temperatures observed by the superconducting oxides.
The gap ratio for the
mean and maximum gap ratio shows some universal features which we
believe to be valid for any choice of the anisotropy function. (i)
The maximum gap ratio lies above the BCS value, increases with
increasing anisotropy, and saturates at a constant value. (ii) The
mean gap ratio is lower than the BCS value and decreases with
increasing anisotropy. We regard those properties of an anisotropic pairing
interaction as universal, i.e. independent of the details of the anisotropy
function and the shape of the Fermi surface.
One should also note that the anisotropic pairing accounts for the gap
anisotropy and high values of the gap ratio $2 \Delta / k_B T$ in a very
elementary
way in the framework of a {\it weak} coupling model.
A more detailed experimental analysis of the
gap anisotropy would be needed to test our ideas.
\vfill
\leftline{\bf Acknowledgement: }
We would like to thank Dr.\ D.\ Einzel and Dr.\ M.\ Schaden for
helpful discussions. We acknowledge financial support from
the Deutsche Forschungsgemeinschaft. This work was also supported by
the National Science Foundation, through grant DMR-91-15491 and
through the Harvard Materials Research Laboratory.

\newpage
\narrowtext

\end{narrowtext}

\newpage

\widetext
\figure{ YBaCuO: polardiagram of the gap (solid); anisotropy
function $f(x=\cos \theta)$.}

\phantom{.}

\figure{ The gap ratio ${2 \Delta  / k_B T_c }$ as a function of the gap
anisotropy for $\nu=1.5$ for the maximum gap $\Delta_{\rm max}$ (solid line)
and the mean gap $\Delta_0$ (dashed line).}


\figure{ Temperature dependence of the maximum (solid line) and
minimum (dashed line) gap in YBaCuO.}
\vfill

\end{document}